# Computer Science

Mahyuddin K. M. Nasution [a,*], Rahmat Hidayat [b], Rahmad Syah [c]

[a] *Computer Science, Universitas Sumatera Utara, Jl. Universitas No. 9 Padang Bulan, Medan, 20155, Indonesia*
[b] *Department of Information Technology, Politeknik Negeri Padang, Sumatera Barat, Indonesia*
[c] *Data Science & Computational Intelligence Research Group, Universitas Medan Area, Medan, Sumatera Utara, Indonesia*
*Corresponding author: [\*]mahyuddin@usu.ac.id*

*Abstract*— Possible for science itself, conceptually, to have and will understand differently, let alone science also seen as technology, such as computer science. After all, science and technology are viewpoints diverse by either individual, community, or social. Generally, it depends on socioeconomic capabilities. So it is with computer science has become a phenomenon and fashionable, where based on the stream of documents, various issues arise in either its theory or implementation, adapting different communities, or designing curriculum holds in the education system. Generally, there are gaps between rich and poor, masculine and feminine, and basic sciences, engineering, and social sciences in accessing and using computer science. In particular, there have been various developments in hardware, software, and brain-ware from computer science in the application. There are recurring problems shortly, in the same or different places, both in developed and technologically lagging countries. Although computer science is still new, it has provided various innovations through research, which is proven by many documents. Therefore, following the growth of computer science - documents in the reputable indexed database, Scopus - there are clues for showing the trends. That is, dividing documents into three-time sessions: sloping, increasing and decreasing. That is by using a growth graphic. Then equip it with some parameters: topics and the weights from hit counts of databases. It gives the meaning as an interpretation to support the reviews. Therefore, this paper will provide a commentary related to that growth. It is for showing that despite the spread of studies, at the last time-session, the issues were always present simultaneously at the same time in accumulation. That is for describing a study title of computer science about what the world already has, going, and will understand about computer science. It is a view regarding its philosophy, basic science, applications, and technology as concepts and solutions that make computer science the primary foundation for skills and professions in the 21st century through the education system. Therefore, other scientific fields such as informatics, computer engineering, information science, information technology, and data science, which are at the crossroads of computer science, are also scientific fields that raise different issues.

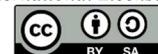

## I. INTRODUCTION

Today, computer science has become a study program in many parts of the world [1], [2]. Computer science also primarily supports many human activities ranging from trivialities to determining the fate of a nation [3]–[5]. With various characters and competencies, computer science has produced many experts [6], [7]. The professionals are at diverse levels for working in different occupations [8]. During the disclosure of computer science simultaneously, various problems arise from computer technology and produce scientific units systematically, each specifically referred to as a unit of lecture material from a curriculum [9], [10].

Conceptually, computer science from the opening comes from understanding and implementing various training materials to carry out an activity with computers [11]–[13]. Specifically, each concept develops into independent parts, then concepts have a relationship with one another through different theories underlying it [14]–[16]. Along the way, study after study related to computers continue to improve the quality and capacity of computers as the tools to complete human welfare [17], and it gradually improves the material of computer science [18], [19], with the evidence attached as literature [20]. However, on the way to becoming a science, computers encounter issues that enhance scientific principles. The phenomenon always meets more than one paradigm in many opportunities for discussion in scientific meetings. This dissemination presents papers in proceedings as scientific publications, while the article provides each problem with a complete solution as scientific findings or innovations [21].

Through literature, dissemination defines it as computer science, where scientific documents reveal its growth and can always dynamically change according to the theory and implementation discussed [22]. This article aims to reveal various problems as a review where there are concepts and solutions so that computer science has become the primary foundation in the 21st century as both a skill and a profession. In material and method: The review starts from the concept that leads to the science or technology contained in the document stream. The next subsection reviews an approach based on the growth of scientific documents. Then result and discussion: It provides an overview of the growth results on several important issues. Finally, reveal computer science in its accumulation and as a spread. All of this is to explain computer science, in general, or particular, and how the world understands that computer science.

## II. MATERIAL AND METHOD

After the appearance of the term "*computer science*" [23], [31] for naming a group of scientific fields related to computers, scientists expressed the science of computers in various ways, mainly to utter objectives in contact with education [32]. However, computer science is a rapidly growing discipline. Computer science has an influence not only on the specialty of the computer field but also on the broader scientific community [33].

*A. Material: From concept to science or technology*

The word "*computer*" is a term that refers to the presence of technology. A noun refers to a tool for ensuring that "*to compute*" carries out the meaning, i.e., to perform computations involving reasoning and thinking. Of course, it is a scientific tool. As a computation tool, computers have proven very useful. Computer science is a science that regards itself as a technological discipline whose purpose is to create problem-solving tools. Useful tools also for the development of other disciplines [34]. Therefore, as a science, it is necessary to have a curriculum planning approach in which computer science is a presentation to all stakeholders [23], [25]. Of course, it requires a structured field guide as an expert. For that reason, change becomes a key feature of situations requiring keeping contact with other disciplines [35]. It aims to make computer science remain stable into science and produce experts following the growth needs. Thus, computer science was a new academic discipline established to serve the professional development needs of the computing community [36]. One of the contacts between science fields is the interpenetration of computer science and language science (linguistics) [37].

As technology, computers have become a phenomenon and the people who use them understand well its necessity. Technology is not only the hardware but also the programmed and living machines (software), something that can adapt to the mind. Computer science is the study of phenomena around the computer [38]. Every type or generation of a computer requires a programming language with which humans communicate to it. On the one hand, it results in the choice of the programming language as a pedagogical aid in teaching computer science, and on the other, it requires sustainability studies. It is when FORTRAN can no longer meet the demands of computer machines and users. Rebellion against FORTRAN has spawned different languages such as PL/1, ALGOL, Pascal, and others [39]. A programming language is a communication tool with computer machines. A computer machine is still a machine. Therefore, from a computer science point of view, there is a chance of interpretation from the computer machine to the machine language. Next, there is a shift from the assembly language to the programming language. All require an interdisciplinary to understand the internal activities of computers. Each interdisciplinary has the characteristics, the experience of administration and operation of the program [40], which allows for the birth of other new programming languages, so a new science again.

Thus, early reviews of conceptual about computer science have revealed the role of computer science through the following definitions [40,41]: Computer science is

(D1) "*the study of computer-related phenomena*" [42],
(D2) "*the study of algorithms*" [43],
(D3) "*the study of the structure of information*" [44], or
(D4) "*the study and management of complexity*" [45].

Therefore, at that time, computer science does not yet have a standard about what, how, and about what level of teaching achievement it should be [46], including considering the need for other fields such as social [47]. Conversely, through its systematic principle, computer science is suitable for guiding error-free design, and it includes designing hardware that supports modern programming concepts [48].

Empirically, computer science has a relationship with natural sciences and even has roots in mathematics, even though related technology plays a more significant role in it [49]. For example, experiments relating to the computer science literature verify Lotka's law (K1) but not the other fields [40]. Thus, interest in computer science originated in various scientific environments and led to various computer science departments with different names appearing [51]. However, at the beginning of the growth of computer science, it still faced obstacles in teaching to operate computers even though various languages adapted to the environment [52]. The deficiency triggered by the axiomatic basis of computer science based on arithmetic theory is weak [53]. One of the approaches offered is to choose a teaching module that suits the learner's interests and background [54].

In science, including computer science, the rejuvenating theory requires experimentation when it has grown up and tends to break down [55]. The implementation of these experiments requires preparation and completeness. Also, to redefine the scientific fields of computer science as is experimental computer science as a non-theoretical activity [56]. The change in computational models is an aftermath of shifting from large computers to small computers (microcomputers). It shows that there are weaknesses in computer science that are not the same as the physical universe model [57]. It facilitates the implementation of the curriculum. For example, in the initial state, it involved interactively teaching programming via BASIC and its interpreter [58]. In this case, it includes preparing teaching materials at lower education program levels before entering college with the convenience available [59]. Then, for example, prepare for synchronization at a higher level by honing numerical skills through FORTRAN in microcomputers for engineering studies program [60]. Unfortunately, technology will always give gaps between communities, cities and villages, especially between countries with different economic capacities. Global issues are already isolated, and the need to engage all parties so that the world accepts computer science [61].

Until the 1980s, the teaching emphasis of computer science in many curricula was to operate computers properly. However, it had not yet revealed the theories and philosophies that systematized those curricula to produce applications, as illustrated by the following two definitions [62]: Computer science is

(D5) "*the study of data, information and knowledge at different levels of complexity or abstraction, and from which other fields of scientific discipline can be developed*" [63], or (D6) "*the technology that deals with the development and use of certain types of human-created artifacts*" [64].

One of the courses in the curricula deals with the foundations of computer science. That is a course that introduces computers to everyone [30]. Including those who are not part of computer science [65], for example, in programming, by trying to use the LOGO language to perfect language approach BASIC [66]. Likewise, the KILANG language is a substitute composition for programming in Indonesian [67]. Anyhow, the civilization gaps have caused not only the computer science department to grow more and more but also to experience a crisis as a result of the imbalance of education and research between related units scattered across countries [29], [68]. The first indication shows that research in computer science is less common than in other disciplines, although scientific research in mathematics and philosophy relies heavily on non-experimentation. The experimental research in computer science is still inadequate [69]. This indication summarizes the development of computer science in China from 1956 to 1985, which asked for technology transfer from the Western world [70]. Hence, there are proposals to refine computer science by expressing its axioms as they apply to physics [71]. In particular, the emphasis on programming skills and the slight neglect of problem-solving in introductory computer science cause the necessity of an extension of the concept from the simple to the more complex form [72], but this requires more abstraction or something of a theoretical nature. Although logic strengthens computer science in general, programming in particular, computer science has many new problems, including the problem P = NP, which relays the discussion based on mathematics [73]. It bases on the perspective emerging from all disciplines that mathematics is a science for building models of computer study [74].

The trend of growing literature on computer science has shown its independence as a science, Fig. 1. Scientific fields that have a distance from computer science or are not directly related to it dodge and join other science; compare Fig. 1 and Fig. 2. Among them are telecommunications, electrical engineering, management, and industry [75], or they are grouped into other fields such as software engineering [27], [76], information technology [77], computational science [78], by disclosing specific materials such as data modeling [79] and the internet [80].

Thus, the information streaming in the literature has led to the birth of the different concepts about computer science. From its inception as an idea until its implementation in the laboratory [81], computer science studies abstract computing rather than concrete mechanisms that perform computations [82]. As a case, computer science is also an academic discipline with roots across disciplines [83], with an instructional design [84] and involves a language approach to the facts at hand [85]. A concept that expresses the importance of a theory being the foundation or it integrated into the curriculum formally from the beginning, such as mathematics [86]. While in the sense of physics and biology, computer science is an experimental science [87], which is one of the dimensions of the computational discipline [88]. It is a science with the ability to analyze itself in its knowledge, is computer science where it is compared with other sciences [89]. It enables computer science as the fundamental of creative endeavors [90], forming a rapidly changing field in which academic [91], vocational [92], and professionals [93] become important and develop into issues related to it [94]. In this sense, computer science is a real phenomenon, where domestic issues clash with the open world [95], and it requires that education involves projects suitable for teaching [96], [97], and projects that robbing it into professional practice in the context of educational research [98].

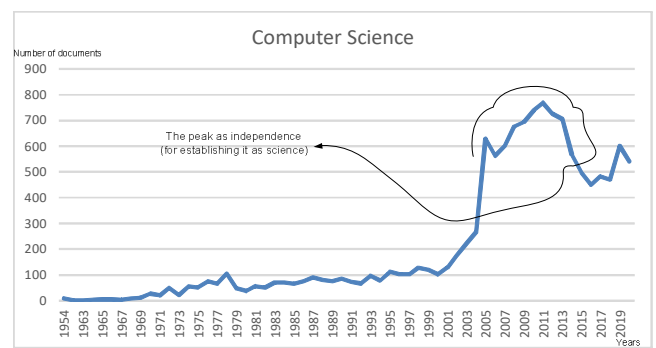

Fig. 1 Literature stream with computer science titles on the Scopus-reputed database 1954-2020.

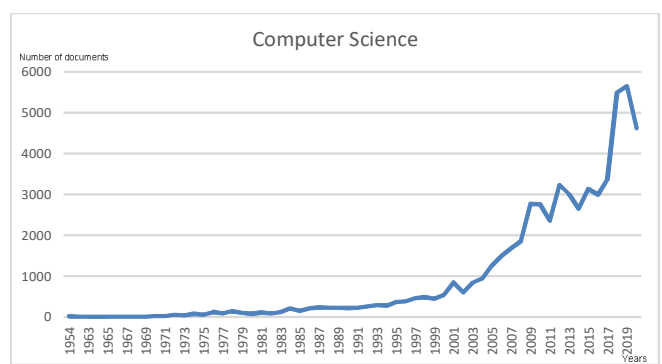

Fig. 2 Literature stream with computer science in documents abstract from the Scopus-reputed database 1954-2020.

Based on research, computer science is a field with knowledge acquisition as a bridge with the concept [99], whereby research transforms computer science rapidly to adapt to complex issues [100], especially challenges to the implementation of reason and logic [101]. While from the public service point, computer science is a discipline that gives rise to paradigms. That is a new tool for all levels of education programs, and it also is to serve the interests of society [102].

However, in the early years since the term computer science accompanied computer technology until the year 2000, it has revealed various intersecting problems between the scientific parts of computer science. Therefore, there are various concepts to define computer science. They related to basic science D1, software D2, hardware D4, functions, and goals of computer science D3-D5-D6. The concepts are to bridge the gap (K1) created by technology until economic-business and socio-cultural communities provide general ideas of solutions.

*B. Method: Approach to growth*

Based on the literature, the information stream about (the concept to the theory of) computer science has revealed problems. The main problem emerges from the cultural disparity in which technology plays a role [103]. Since the beginning, computers as technology is something have been fashionable [104]. Computers, both academically and in business, are all-around tools. It seems that computer science is a scientific field that has surprises. Research supported by basic science such as mathematics, physics, and electronic engineering has spawned many innovations for and around computers. As with Moore's law [105], much of that technology becomes obsolete because of other inventions, or outdated technology (K1) becomes the everyday support of a culturally uninspired world to make technological leaps [106]. The development of computer science based on evidence is as publication grew sloping until 2000. Call it $x_1$. Then the evidence rose sharply until its peak in 2010. Call it $x_2$. After that decreased until 2020 or $x_3$; see Fig 1. An overview of some things related to computer science will present as follows. It will support by tables of scientific publications about discussion topics.

By proposing a paradigm, the approach for completing the main problem based on true methods is theoretically a way to research computer science. Research affects computer science and basic science such as mathematics, physics, and biology, from which many fields of computer studies developed [107]. On the brain-ware side, a research approach to computer science has proposed to bridge gender differences in computer science [108] and reveal differences in performance [109]. Research and education collaboration is an approach from the human resource development perspective, intending to elucidate the difficulties in elaborating the computer science curriculum [110]. It aims to bring the attention of all parties by accelerating technology transfer as a refresher. Of course, they are novice learners of science. Another proposal supports technology transfer using graph theory, animation, and the web, where remote programming with minimal hardware requirements is an approach to mastering computer science [111]. The development of hardware, however, did not come from the same factories, so it presented a different platform, which begged the operating systems are not the same to operate. The presence of an open-source operating system as an approach to technology transfer also presents different languages. It requires uniformity, if not standardization, namely non-platform languages as a solution, such as Java [112]. However, the languages are generally derived from previous procedures such as Pascal and C [113].

Transfer of technology can involve the computer itself as a model of exploration. That is a virtual museum that performs features of definite programming languages interactively and self-taught. Thus, it enables integration with the curriculum [114]. However, this technology transfer is inseparable from a new wave of theories and tools working for data. In this case, research reveals the importance of theory, methods, models, and systems for data studies such as statistics [115].

The approach to various concepts is to incorporate definitions of computer science and computational disciplines: Computer science is the science of information processes and their interactions with the world through the support of systematic studies by which algorithms describe and transform information, theory, analysis, design, efficiency, effectiveness, optimization, implementation, and applications [116]. In short, computer science is a science that involves computing. However, a definition of computer science is not sufficient to limit the growth of its fields. Computer scientists not only disagree with computer science as a science, but they also disagree with the discipline's content, form, and practice. Therefore, the definition of computer science flows according to need. Computer science is a broad and rapidly growing discipline that allows knowledge creation, maintenance, and modification, but the latter is often not part of computer science [117]. Therefore, it affects the interests of prospective scientists in a gendered manner [118]. However, formally computer science is a science that firmly has a theory and its implementation in computer programs with rigid language and standards and has its metaphor. Computer science provides a conceptual framework by placing the disciplinary ontology of computers in a computing environment [119].

III. RESULT AND DISCUSSION

The terminology and discourse of computer science follow the facts that are full of metaphors. It has to do with widespread human interaction and technology in various languages [120]. However, the awareness of globalization and technological equity follows the flow of information, especially in the literature [121].

*A. Results of growth*

Several explanations reveal the journey of computer science phenomena and paradigms by involving available technologies – such as search engines, databases, queries, and taxonomies. It starts from a scientific basis.

*1) Scientific basis*

Computer science has a scientific basis as a source of knowledge, i.e., mathematics. Mathematics and philosophy are two co-existing knowledge, which can be said to be twins [122]. Through dialogue, philosophy tries to justify it, while mathematics elaborates and proves the point. Therefore, the philosophy of computer science did not thrive [123]. On the other hand, computer science and mathematics are closely interdependent because it is necessary, but it does not suffice to understand and solve the existing problem [124]. However, based on the philosophy of the knowledge generalization, i.e., ontology [125], the existence of computer science conceptually, through the query "*computer science is*" against document abstract, and by querying "*computer science*" in the title against Scopus reputable indexer databases reveal a picture of the development trend of computer science [126]. This review is further based on data from the literature and an overview of computer science based on the growth of concepts in recent years [127]. That is, based on Table I, there was a spread of studies when the number of documents increased sharply, but the focus of mathematics-related studies was quite high.

TABLE I
PHILOSOPHY AND MATHEMATICS FOR COMPUTER SCIENCE

| Topics | Number of Documents | | | |
|---|---|---|---|---|
| | $x_1$ | $x_2$ | $x_3$ | Total |
| Philosophy in title | 6 (24.00%) | 5 (20.00%) | 14 (56.00%) | 25 |
| Philosophy in abstract | 36 (38.71%) | 17 (18.28%) | 40 (43.01%) | 93 |
| Mathematics in title | 72 (31.86%) | 49 (21.68%) | 105 (46.46%) | 226 |
| Mathematics in abstract | 108 (25.71%) | 98 (23.33%) | 214 (50.95%) | 420 |

*2) The components of a computer*

Computer science is a relatively young discipline compared to other science. A scientific field that combines science, engineering, and mathematics as is knowledge, skill, and a scientific basis [128]. Thus, since the very beginning, both theory and implementation of computer science have involved issues related to hardware, software, brain-ware, and data.

Mathematically in discrete mathematics [129], some theories encourage the birth of innovations in hardware ranging from logic [130], Boolean algebra [131], graph theory [132], and others [133]. Challenges in hardware development come from electronic systems that require engineering. It indirectly requires hardware laboratories to carry out experiments. In this case, discrete mathematics acts as the scientific basis of the physically present technology or hardware [134]. Thus, the features of computer hardware are the technology. It is since the beginning provides modules that can be presented part by part. Each module keeps adapting to the interests of the use and the increasing capability of the technology [135]; see Table II about hardware and machine.

The syntax and semantics of natural languages form the basis for designing and building programming languages [136]. Logically, the programming language has a beginning and an end in implementing logic as a sequence of commands for the computer to carry out [137]. At first, computers only understood machine language, then assembly languages were born, and then higher-level languages [138]. Machine language and assembly languages build the operating system [139]. Then higher-level languages build the more complex systems [140]. The implication of the symbolic and systematic arrangement of symbols is software, including programming languages and systems running on the computer [141]. Mathematically, the software requires a scientific field from arithmetic that underlies calculation principles [142]: permutation, enumeration in set theory, relations and functions, language and finite state machines, number systems, principles of inclusion and exclusion, linear algebra or algebra, group theory to ring theory and modules, coding theory, combinatorial design, graph theory to network analysis, trees and optimization [143]–[145]. The topics of software and algorithm in Table II support the above description, where the software component is the hottest issue.

TABLE II
RELATED COMPONENTS OF COMPUTER SCIENCE

| Topics | Number of Documents | | | |
|---|---|---|---|---|
| | $x_1$ | $x_2$ | $x_3$ | Total |
| Hardware in title | 13 (50.00%) | 6 (23.08%) | 7 (26.92%) | 26 |
| Hardware in abstract | 79 (30.86%) | 58 (22.08%) | 119 (46.48%) | 256 |
| Machine in title | 11 (30.56%) | 8 (22.22%) | 17 (47.22%) | 36 |
| Machine in abstract | 97 (23.26%) | 86 (20.62%) | 234 (56.12%) | 417 |
| Software in title | 106 (35.33%) | 91 (30.33%) | 103 (34.33%) | 300 |
| Software in abstract | 288 (24.57%) | 343 (29.27%) | 541 (46.16%) | 1172 |
| Algorithm in title | 20 (28.57%) | 12 (17.14%) | 38 (54.29%) | 70 |
| Algorithm in abstract | 207 (18.65%) | 278 (25.05%) | 625 (56.31%) | 1110 |
| Operator in title | 1 (33.33%) | 0 (0%) | 2 (66.67%) | 3 |
| Operator in abstract | 12 (16.90%) | 18 (25.35%) | 41 (57.75%) | 71 |
| Administrator in title | 0 (0%) | 1 (50.00%) | 1 (50.00%) | 2 |
| Administrator in abstract | 5 (12.50%) | 18 (12.50%) | 41 (50.00%) | 40 |
| Programmer in title | 3 (60.00%) | 1 (20.00%) | 1 (20.00%) | 5 |
| Programmer in abstract | 22 (24.66%) | 22 (24.66%) | 49 (52.69%) | 93 |
| Experts in title | 2 (15.38%) | 4 (30.77%) | 7 (53.85%) | 13 |
| Experts in abstract | 21 (13.04%) | 32 (19.88%) | 108 (67.08%) | 161 |
| Data in title | 23 (17.16%) | 27 (20.15%) | 84 (62.69%) | 134 |
| Data in abstract | 230 (13.04%) | 309 (19.88%) | 1031 (67.08%) | 1570 |
| Information in title | 68 (14.56%) | 164 (35.12%) | 235 (50.32%) | 467 |
| Information in abstract | 194 (15.30%) | 324 (25.55%) | 750 (59.15%) | 161 |
| Knowledge in title | 3 (15.38%) | 24 (30.77%) | 74 (53.85%) | 101 |
| Knowledge in abstract | 123 (12.15%) | 226 (22.33%) | 663 (65.51%) | 1012 |

The term "brain-ware" does not recognize in computer science [146]. A term refers to human resources. The people are involved in developing science, technology, and applying computers [147]. Of course, everyone's involvement in the use of technology is after the birth of technology [148], but mastery of science starts in the individual mind and accelerates through teaching and learning from school to college [149]: undergraduate and postgraduate [150,151]. In addition to the interaction between learners and teachers, the accelerating human resource development for mastering the fields of computer science may involve online courses and laboratories [152,153]. Human resources are one component of computer science, which is why computer science produces technology, such as the Web, to carry out online learning, including independent computer science learning [154]. Therefore, the assumptions about brain-ware relate to operators, administrators, programmers, experts, etc. See Table II.

The core goal of computer science is about data where processing is urgent [155], see Table II. Because it is computation makes it possible to present information [156], visualizations [157], and sounds such as music [158], where data acquisition plays a role in measuring performance [159]. Based on data as the target of computer science, computer technology encourages the generation of parts of research from computer science that touch various aspects of life, such as bioinformatics [155], [156], [159], [160].

*3) Scientific publication*

As previously stated, computer science is a discipline that continues to develop and even breaks down into more centralized scientific collections. That is study program be more centralized according to the original concept of each [161], [162]. It is shown by the literature growth data entitled computer science, and it is directly different from the existence of computer science in the abstract of literature.

TABLE III
SCIENTIFIC PUBLICATION: PROCEEDINGS OR JOURNALS

| Topics | Number of Documents | | | |
|---|---|---|---|---|
| | $x_1$ | $x_2$ | $x_3$ | Total |
| Publications in title | 3 (6.98%) | 10 (23.26%) | 30 (69.77%) | 43 |
| Publications in abstract | 18 (7.03%) | 61 (23.83%) | 177 (69.14%) | 256 |
| Proceedings in title | 97 (9.75%) | 328 (32.96%) | 570 (57.29%) | 995 |
| Proceedings in abstract | 243 (15.50%) | 450 (28.70%) | 875 (55.80%) | 1568 |
| Journals in title | 26 (21.49%) | 60 (49.59%) | 35 (28.93%) | 121 |
| Journals in abstract | 18 (10.53%) | 46 (26.90%) | 107 (62.57%) | 171 |

Computer science is a science in which proceedings are the primary publication site [163], [164]. That is the dissemination through scientific events and activities. They are seminars, conferences, symposia, etc. Apart from the purpose of exchanging scientific information, they are primarily to claim an innovative draft. A paper is a simple scientific publication, usually with one objective. An article is a complete scientific publication with at least two objectives. Thus, papers can convey the latest study information and discuss it in scientific meetings.

Meanwhile, other authors cite the articles in papers, and the quotations mostly from papers in proceedings with other objectives are disclosures towards new problem statements [165]. Many computer science scientific activities are dedicated to specific themes, reaching out to a more specialized scientific community so that specific research can develop into something new or have a new perspective [166]. Thus, the scope of studies in journals is getting wider, or the emergence of new journals to accommodate articles within their scope [167]. Of course, it indirectly increases innovation on all sides [162,168]. It changes not only users but also production behavior [169]. For example, the production of scientific papers must have standards ranging from categorization, publishing, indexing, and ranking [163], [165], [166], [168], [170]. Therefore, computer science is a science that continues to grow through evidence of research, i.e., scientific publication [171], see Table III.

*4) The courses until curricula*

Computer science is a growing discipline that periodically itself [126]. This growth is related to institutional changes in the curricula of the study programs. It is the pressure from theoretical and applied scientific fields of computer science also [172]. The role of the bioinformatics discipline, for example, is to reveal information about the behavior of living cells in biology. Revealing information through the involvement of other sciences, for example, it is the probabilistic theory [173]. It is related to also the appearance of the discipline of decision-making is the idea of increasing the effectiveness of economic activities through systems involving computer networks and the internet [174]. Thus, curriculum design depends on the directional pressure of competencies. See Table IV.

TABLE IV
CURRICULUM FOR PROGRAM STUDY OF COMPUTER SCIENCE

| Topics | Number of Documents | | | |
|---|---|---|---|---|
| | $x_1$ | $x_2$ | $x_3$ | Total |
| Curricula in title | 243 (45.94%) | 126 (23.82%) | 160 (30.25%) | 529 |
| Curricula in abstract | 322 (25.91%) | 334 (26.87%) | 587 (47.22%) | 1243 |
| Program in title | 170 (41.77%) | 91 (22.36%) | 146 (35.87%) | 995 |
| Program in abstract | 472 (28.35%) | 428 (25.71%) | 765 (45.95%) | 1568 |

Even though computer science is already well-established in higher education [175], an introduction to computer science is the opening course in each curriculum [176]. That connects insights that are ready and accepted at the education level below with other scientific units [177], especially computer science, in universities. Introducing and maintaining computer science in many schools as a subject is a consideration for technology transfer [175]. Scientific fields in computer science continue to grow and update their content. The two last activities are according to the demands and competitiveness of alumni. There are new areas of study such as Grid Computing, Cyber-security, Robotics [126], and specific courses like the Quantum Computing scientific units. Quantum computing offers a new paradigm to computer science so that computer technology with Turing machine properties has more storage capacity, solves artificial intelligence well, and computers perform higher processing and more [178]. However, regularly reaffirming computer science, each computer science curriculum is mandatory for the following improvements. By using the research outputs [179], the birth of new scientific units [180], the local competence [181], and the technological capability of a country [182] are among the issues for complementing the improvement [183]. Changes in the computer science curriculum also allow changes to other programs such as computer engineering [184], informatics [185], information science [186], information systems [187], software engineering [188], etc.

### 5) Masculine

Computer science is a science that looks gendered and masculine [102], [189]. From the early days of computers, there have been studies of different gender interests in computer science [190], [191]. Computer science is both theoretical and technical. Perhaps, the theory has put pressure on humans to think hard, especially in implementing logic in programming [192], while techniques reveal that humans have to work hard [189]. It has resulted in women's general reluctance to study or work on computers [193], but elsewhere computer science has attracted much interest from women [194]. Of course, there are attractive factors for women, and perhaps things like the graphic and animation to the design give influence the presence of that interest. Based on these studies, in computer science, brain-ware or human resources have specifications based on gender [195], [196], see Table V.

TABLE V
GENDER ON COMPUTER SCIENCE

| Topics | Number of Documents | | | |
|---|---|---|---|---|
| | $x_1$ | $x_2$ | $x_3$ | Total |
| Gender in title | 9 (6.98%) | 38 (29.46%) | 82 (63.57%) | 129 |
| Gender in abstract | 21 (7.45%) | 61 (21.63%) | 200 (70.92%) | 282 |
| Masculine in title | 0 (0%) | 1 (100.00%) | 0 (0%) | 1 |
| Masculine in abstract | 1 (7.69%) | 6 (46.15%) | 6 (46.15%) | 13 |
| Women in title | 27 (22.13%) | 39 (31.97%) | 56 (45.90%) | 122 |
| Women in abstract | 30 (10.71%) | 85 (30.36%) | 165 (58.93%) | 280 |

The involvement of women in teaching and learning in particular, and generally the development of computer science, is related to culture, socio-economy, and prosperity [190]. Different education systems between countries or regions influence women's interest in computer science [194]. Therefore, each of them thus requires its strategy [197], for example, relying on a country's regional potential or strength, such as relying on the game as a lure [198].

### 6) Education system

Computer science is a science that affects the education system [150]. That is a science that introduces new rules to the academic communities for designing, planning, developing, customizing, and implementing curriculum [199]. The curricula range from hardware assembly and software installation, operating system administration, and programming to an application [147]. Therefore, students' interest in computer science is growing as well as the development of its implementation in managing data and financial worksheets [153], [195], in addition to building good collaboration between students through networks and games [151] and robotics [157], also developing engagement between institutions [200,201].

The presence of the Web has made it possible for all types of media to be something integrated, apart from being a means of two-way scientific communication [202], [203]. Types of media are multimedia. Types of media are multimedia. And then delivery of teaching materials and questions and answers to be easily carried out, as well as being self-taught and able to be studied and repeated by learners.

TABLE VI
EDUCATION ON COMPUTER SCIENCE

| Topics | Number of Documents | | | |
|---|---|---|---|---|
| | $x_1$ | $x_2$ | $x_3$ | Total |
| Education in title | 278 (25.39%) | 252 (23.01%) | 565 (51.60%) | 1095 |
| Education in abstract | 275 (15.27%) | 385 (21.38%) | 1141 (63.35%) | 1801 |
| Academic in title | 13 (17.33%) | 14 (18.67%) | 48 (64.00%) | 75 |
| Academic in abstract | 107 (17.20%) | 135 (21.70%) | 380 (61.09%) | 622 |
| Vocation in title | 1 (16.67%) | 2 (33.33%) | 3 (50.00%) | 6 |
| Vocation in abstract | 5 (16.67%) | 7 (23.33%) | 18 (60.00%) | 30 |
| Professional in title | 18 (18.95%) | 11 (11.58%) | 66 (69.47%) | 95 |
| Professional in abstract | 82 (15.02%) | 122 (22.34%) | 342 (62.64%) | 546 |

It is the purpose of the changes in the education system [204]. Computer science study programs have always been at the forefront of educational system changes [205]. Facilities and infrastructures directly cause many computer scientists to become good teachers [206]. They are good teachers for teaching or computer science, but also for other areas of their expertise. The educational components in a system, such as learners, teachers, teaching materials, facilities, and infrastructure, are integrated with and by computer science [207]. While the implementation of teaching and learning of computer science in the education system involves different approaches: academic, vocational, and professional (training), see Table VI.

### B. Discussion: Accumulation and spread - the growth overview

In concept and theory, then to application, the computer science literature reveals consistent issues. It flows with the growth of documents in response to the changing challenges that arise from every activity of human life over time. Philosophically, computer science is a science with issues that arise based on reflecting on the discipline and its scientific practice [208]. In the computational artifacts of the philosophy of computer science [209], computer science is a science that involves logic to realize the systematics of ordinance and mathematics for modeling as the basis for developing the theory [210]. Although philosophy did not underlie the emergence of computer science [122], this requires a collaborative transformation between philosophy and computer science that presents an interaction between teaching and teaching [211]. Computer science is a science that provides technicality also theory to solve issues of life. There is the paradigm for phenomena through mathematical modeling and its interpretation by programming to produce something that lives in a computer, namely software [212].

Mathematics is a language for understanding science and technology: computer science and itself computer [210]. However, computer science is a science that causes a paradigm shift from classical natural science to the context of

1148

computational science [213]. It causes computer science to be a dynamic and highly innovative science, i.e., something that causes the user community to be a digital society where its products significantly contribute to the development of digital science and technology [214]. Creativity demands from today's industrial world: Something relevant is that computer science is a creative field to work in where the software design process is a testament to an effort in that direction [215]. In particular, computer science is a science that encourages innovation and productivity in societies that enrich with technology in their work and education [216] by creating something new. That is creativity in various fields such as coding, open-source projects, video games, augmented reality, and artificial intelligence [217], creating many kinds of software, web applications, mobile applications, and others [218]. However, computer science is an established science with a natural selection of technologies where various technologies are rapidly becoming obsolete [219].

From an educational point of view, computer science is a science that has grown so important to society since its inception [220], but its growth has several factors to be resistances. The main interests and resistances have a root in human nature where it negatively affects the world of education: Political influence [121]. The others are gender issues [221] and technology social influences [222].

In the political world using it, computer science is a science that brings technology to society to destroy social orders by reducing face-to-face interactions [223]. However, ethics and scientific principles alleviate resistances where computer science is a practical discipline that evaluates learners' practices to address alleged scientific violations such as plagiarism [224]. Computer science is a science that reduces its resistances. It has been done by determining the universalization of access to science by using equity as an agreement in the educational community [225]. In this case, there is the forming of an identity of abilities such as institutions as a form of recognition [226], creating a structural, instructional, and curricular learning ecosystem for programs up to partnerships with local organizations, namely a community that forms an ecosystem based on equity [227]. Thus, computer science is a science with a core discipline in education through defining and formulating scientific goals [228]. Therefore, computer science is a science that has given birth to scientists with the most desirable careers in the world [229].

Although computer science is a science with some masculine scientific fields [230], [231], [232], even so, there are still fields of computer science that are not attractive to most people [233]. However, providing basic knowledge about computer science early, that is, by introducing it to lower levels of education, gives better hope for mastering the knowledge [234]. Introducing computer science in schools [235], implementing a week's workshop for the beginners, or they follow matriculation of computer science [236]–[238], are the conceptual bridge that requires related knowledge but transforms it also [239]. It is to convey strong reasons to study scientific fields related to computer science [240].

Furthermore, computer science is a science with various strategies to reduce issues that present in diverse events. That is by developing relevance to increase the number of enthusiasts [241] or by feeding ideas or theories from computer science into the mathematics Olympiad [242]. A way is about revealing the relationship between computer science and games related to, respectively, men and women [243], [244] by building equity in employment [245] or by framing performance and interest [246] to address inequity [247], especially in masculine software development [248]. It is also a way of using cohesive interventions regarding socio-ecological psychological systems [249]. Therefore, computer science has a broad ecosystem, where the transformation of science and technology is bilingual [250]. It requires more teaching skills and abilities [251], namely science by developing models of information delivery [252], building an approach by assessing the weakest areas of the learner [253], and analyzing content and pedagogical oncoming in a pilot project [254], [255]. Then, computer science is a science that includes technology as an innovation in teaching [256] by incorporating real-world problems into the classroom [257] and incorporating technical skills and professional quality from the industry [258] through increasing the number of lectures and proficient teacher [259]. Therefore, as a consequence, computer science is a science that has developed very rapidly in the last two decades affecting all areas of human activity, including education, with a variety of curricula and degrees and within different faculties [260]. It generates intensive motivation for forming and socializing it through program development and interfacing various devices and robotics [261], [262].

Computer science, based on mathematics as a reasoning language, describes objects precisely through their relationships. It directly through logic has interpretation in the machine language of computers [263]. Logically, computer science is a science infused with metaphors and influences derived from the interactions of inter-lingual computer users and with the support of decision-making locally and globally [264]. With the support of research on the renewal of computers, technology continues, and computer science is the science that presents many new technologies in the field of telecommunications where this science provides the main challenge for forensics [265], namely science with the support of various paradigms and methodologies [266]. In this case, computer science is a science that combines literacy, theory, and experimentation in teaching [267]–[269]. That is a science that can trace back literature or text documents to reveal human behavior and activity tendencies [270]. It includes seeing the historical roots of research in computer science [271]. Systematic literature studies collect and synthesize empirical evidence through study and machine learning [272]. Computer science requires literacy about a professional career [273]. The output of research related to computer science is in the form of documents or scientific publications. The documents are a quantitative and qualitative scientific practice and show a hybrid development trend [274]. In other words, computer science is a science that starts from the most basic level [275] to literacy of computation in other fields [276]: biology, economics, society, and culture [277, 278], for example, including there is a transfer to other disciplines such as from computer science to geography as a digital revolution [279]. With research output, computer science underlies scientific publication metrics such as the h-index based on the number of citations from Scopus or the Web of Science (WoS) [280]. That underlies the computer science rank matrix as a study program that aims to extract, mine and rank academic information [281]. Thus, computer science is a science that keeps disclosing the output of its research through conferences and workshops. The scientific publications prove its growth rate compared to other fields [282]. The growth is not limited to the computer itself [283].

However, computer science is a science with disciplinary demands both in the profession and in products and uses in various functions [284,285], including differences in the level of knowledge of academics and society so that it demands collaboration [286]. In public service using technology, computer science is a science that produces beneficial outcomes, almost always [287]. Either simple or complex, but easy to use: Starting from creating a user-friendly interface [261]. By providing students with the keys to understanding science: data structures, algorithms, and visualization principles [288]. Establish convenience for software and industrial development, such as trial-driven development (TDD) [289]. In other cases, by developing convenient methods such as drag-and-drop in object-oriented programming [290], involving the cellular/smartphone technology that almost every person has [171], presenting them in-game leagues [291], by following in-game tours [245], or also by involving in the form of adventures [184] where all those are triggering motivation for provoking participation [292]. Therefore, computer science is widespread and always present in everyday life [293]. With its technological potential, computer science has become a science that reduces the cost of life on all sides, including the cost of education [294].

Computer science is a science that makes it possible to explore all aspects of human life [295,296]. Although it is a relatively young science, computer science is a well-established science in education [297]. This science has grown diverse and varied as a result of intense competition [298]. The science forces the presence of curriculum variations [299] to respond to challenges of each learning environment and its different applications [300] with broader and more thorough knowledge [301], but as a whole, articulate a theoretical framework that accommodates programming independently of mathematics [302]. Through it growth, computer science is a science with several features [303], and a science that gives birth to practical skills in the 21$^{st}$ century [304], where there is an advanced level of knowledge that makes studies possible [305]. Thus, computer science is a science that opens up the intersection of disciplines to different sciences but has a close relationship where multi-disciplines have a focal point in computer science makes it something important to others [306]. In particular, computer science is evolving into *Informatics* [307], a scientific unit that supports high-performance organizations through computing [308].

On the other hand, computer science accumulation into scientific units called *Information Science* results from constructive challenges in science and technology with a hybrid methodology both from a content point of view and a context point of view [309]. Certain parts of computer science concentrate on Information Technology, naming groups of scientific fields that shift power to be increasingly centralized organizationally [310]. At the same intersection, computer science is present in *Computer Engineering* programs. Partly adapted to computer technology offers different *Quantum Computation* or *Quantum Information* [177] and claims that the computer systems are more powerful than classical computers [311]. Meanwhile, the language of science and technology: mathematics, and computer science, brings forth the *Data Science* [312] as the language for understanding data [313], [314]. In particular, computer science is the only science that supports the industrial revolution entirely [315], including the industry 4.0 [316], [317], involving many recent developments such as the Web that allows searching-finding, retrieving-clustering, ranking, categorizing, and analyzing information [318]. Therefore, through this rapid growth, computer science has become a science and tool for other disciplines that require lifelong learning [319], a science that has a scientific target [320], and which resolves resistances and limitations [321] by providing a scientific introduction because of those changes [322].

Suppose that the mean percentages of all parameters $x_1$, $x_2$, and $x_3$ are $p(x_1)$, $p(x_2)$, and $p(x_3)$, respectively, from Table I to Table VI. The comparison between the three shows that although issues related to computer science spread from the start, the topics of discussion about computer science significantly related to one another. That is $p(x_1) < p(x_2) < p(x_3)$, or 24% < 27% < 49% for title of documents and 19% < 24% < 57% for abstract of documents.

IV. CONCLUSION

Conceptually, computer science was born from various cultural interests, which technologically there are gaps, but integrated by and on the same basis. Throughout the growth of computer science, various issues have been present and have become topics of discussion on many occasions. The issues have been discussed in an integrated and simultaneous manner recently. Alongside computer science has indirectly presented other scientific units such as informatics, computer engineering, information science, information technology, or what has emerged recently is data science. Thus, it indirectly raises new problems that become studies in computer science or other fields.